\documentclass[10pt,conference,dvipsnames]{IEEEtran} 

\usepackage{changepage} 
\usepackage{cite}
\usepackage{tikz}
\usepackage{pgfplots}
\usepackage{array}
\usepackage{booktabs} 
\usepackage[utf8]{inputenc}
\DeclareUnicodeCharacter{FB01}{fi}

\usepackage{mathptmx} 

\usepackage{mathptmx}

\usepackage{amsmath,amssymb,amsfonts}
\usepackage{booktabs}
\usepackage{csquotes}
\usepackage{graphicx}
\usepackage{verbatim}
\usepackage{fancybox}
\usepackage{multirow}
\usepackage{tabularx}
\usepackage{cellspace}

\usepackage{tcolorbox}
\usepackage{enumitem}
\usepackage{assets/templates/iselab}
\usepackage{pgfplots, pgfplotstable}

\usepackage{bchart}
\usepackage{adjustbox}
\usepackage{tabularx}
\usepackage{pgfplots}
\usepackage{listings}
\usepackage{caption}
\usepackage{varwidth}
\DeclareCaptionType{mylisting}[Listing][List of My Listings] 

\usepackage{color} 
\usepackage{soul} 

\usepackage{cleveref}
\newcommand{\RedL}{\makebox[0pt][l]{\color{red!6}\rule[-3pt]{\linewidth}{10pt}}}
\newcommand{\GreenL}{\makebox[0pt][l]{\color{green!6}\rule[-3pt]{\linewidth}{10pt}}}

\usepackage{mdframed}

\definecolor{orange}{RGB}{255,127,0}
\definecolor{green}{RGB}{0,127,0}
\definecolor{white}{rgb}{0.99,0.99,0.99}
\definecolor{dkgreen}{rgb}{0,0.6,0}
\definecolor{dred}{rgb}{0.545,0,0}
\definecolor{dblue}{rgb}{0,0,0.545}
\definecolor{lgrey}{rgb}{255,0.9,0.9}
\definecolor{gray}{rgb}{0.4,0.4,0.4}
\definecolor{dgray}{rgb}{0.2,0.2,0.2}
\definecolor{darkblue}{rgb}{0.0,0.0,0.6}

\lstdefinelanguage{diff}{
  basicstyle=\scriptsize\ttfamily \color{black},
  columns=fullflexible,
  breaklines=true,
  breakatwhitespace=false,
  showspaces=false,               
  showstringspaces=false,  
  frame=single, 
  showtabs=false,
  numbersep=5pt,
  showstringspaces=false,        
  stepnumber=1,                   
  tabsize=5,                     
  title=\lstname,  
  numbers=left,                 
  numbersep=5pt,   
  backgroundcolor=\color{white},
  morecomment=[f][\lstbg{red!5}]-,
  morecomment=[f][\lstbg{green!5}]+,
  morecomment=[f][\textit]{@@},
}

\lstdefinelanguage{SQL}{
  basicstyle=\tiny\ttfamily\color{black},
  columns=fullflexible,
  breaklines=true,
  breakatwhitespace=true,
  showspaces=false,
  showstringspaces=false,
  frame=single,
  showtabs=false,
  numbersep=5pt,
  stepnumber=1,
  tabsize=4,
  title=\lstname,
  numbers=left,
  numbersep=5pt,
  backgroundcolor=\color{white},
  morecomment=[l]{--},
  morecomment=[s]{/*}{*/},
  morestring=[b]', 
  morekeywords={SELECT, FROM, WHERE, GROUP BY, HAVING, ORDER BY, ASC, DESC, INSERT INTO, VALUES, UPDATE, SET, DELETE FROM, CREATE TABLE, ALTER TABLE, DROP TABLE, PRIMARY KEY, FOREIGN KEY, REFERENCES, AND, OR},
  keywordstyle=\color{green},
  identifierstyle=\color{black},
  commentstyle=\color{green!60!black},
  stringstyle=\color{orange},
}

\lstdefinelanguage{cpp}{
      backgroundcolor=\color{white},  
      basicstyle=\footnotesize \ttfamily \color{black} ,   
      breakatwhitespace=false,       
      breaklines=true,
      postbreak=\mbox{\textcolor{black}{$\hookrightarrow$}\space},
      captionpos=b,                   
      commentstyle=\color{gray},   
      deletekeywords={...},          
      escapeinside={\%*}{*)},                  
      frame=single,                  
      language=C++,                
      keywordstyle=\color{dblue},  
      morekeywords={BRIEFDescriptorConfig,string,TiXmlNode,DetectorDescriptorConfigContainer,istringstream,cerr,exit}, 
      identifierstyle=\color{black},
      stringstyle=\color{blue},      
      numbers=right,                 
      numbersep=3pt,                  
      numberstyle=\tiny\color{dgray}, 
      rulecolor=\color{black},        
      showspaces=false,               
      showstringspaces=false,        
      showtabs=false,                
      stepnumber=1,                   
      tabsize=4,                     
      title=\lstname,
    }

\definecolor{codegray}{RGB}{128, 128, 128}
\definecolor{codepurple}{RGB}{88, 86, 214}
\definecolor{backcolour}{RGB}{245, 245, 245}
\definecolor{codegreen}{RGB}{4, 186, 52}
\definecolor{codered}{RGB}{240, 0, 0}
\definecolor{coderedhighlight}{RGB}{252, 192, 194}
\definecolor{codegreenhighlight}{RGB}{192, 252, 198}
\usepackage{pgfplots}
\usepackage{tikz}
\usepackage{array}
\usepackage{booktabs} 

\lstdefinestyle{github}{
    aboveskip=15pt,
    basicstyle=\scriptsize\ttfamily,
    breakatwhitespace=false,
    breaklines=true,
    commentstyle=\color{codegray},
    stringstyle=\color{black},
    frame=tb,
    numbers=left,
    numbersep=3pt,
    keywordstyle=\color{codered}\bfseries,
    numberstyle=\tiny\color{codegray},
    showspaces=false,
    showstringspaces=false,
    showtabs=false,
    tabsize=2,
    moredelim=[l][\color{codered}]{-},
    moredelim=[l][\color{codegreen}]{+}
}

\begin{document}

\title{Build Code Needs Maintenance Too: A Study on Refactoring and Technical Debt in Build Systems}

    

    

\author{\IEEEauthorblockN{Anonymous Authors}}
\author{\IEEEauthorblockN{1\textsuperscript{st} Anwar Ghammam}
\IEEEauthorblockA{\textit{Oakland University} \\
Rochester Hills, USA \\
anwarghammam@oakland.edu}
\and

\IEEEauthorblockN{2\textsuperscript{nd} Dhia Elhaq Rzig}
\IEEEauthorblockA{\textit{University of Michigan- Dearborn} \\
Dearborn, USA \\
dhiarzig@umich.edu}

\and
\IEEEauthorblockN{3\textsuperscript{d} Mohamed Almukhtar}
\IEEEauthorblockA{\textit{University of Michigan- Flint} \\
Flint, USA \\
almukhtr@umich.edu}

\and
\IEEEauthorblockN{4\textsuperscript{th} Rania Khalsi}
\IEEEauthorblockA{\textit{University of Michigan- Flint} \\
Flint, USA \\
rkhalsi@umich.edu}
\and
\IEEEauthorblockN{5\textsuperscript{th} Foyzul Hassan}
\IEEEauthorblockA{\textit{University of Michigan- Dearborn} \\
Dearborn, USA \\
foyzul@umich.edu}

\and
\IEEEauthorblockN{6\textsuperscript{th} Marouane Kessentini}
\IEEEauthorblockA{\textit{Grand Valley State University} \\
Grand Valley, USA \\
kessentm@gvsu.edu}

}
\definecolor{promptcolor}{HTML}{2D4671}
\definecolor{devcolor}{HTML}{805A15}
\newenvironment{promptbox}[1]{%
  \mdfsetup{%
    frametitle={%
      \tikz[baseline=(current bounding box.east),outer sep=0pt]
      \node[anchor=east,rectangle,fill=promptcolor]
      {\strut\footnotesize\textcolor{white}{#1}};},
    innertopmargin=0pt,
    linecolor=promptcolor,
    linewidth=2pt,
    frametitleaboveskip=\dimexpr-\ht\strutbox\relax,
    skipabove=4pt
  }
  \begin{mdframed}[]\relax%
    \scriptsize\ttfamily 
    }{\end{mdframed}}

\newenvironment{devbox}[1]{%
\mdfsetup{%
  frametitle={%
    \tikz[baseline=(current bounding box.east),outer sep=0pt]
    \node[anchor=east,rectangle,fill=devcolor]
    {\strut\footnotesize\textcolor{white}{#1}};},
  innertopmargin=0pt,
  linecolor=devcolor,
  linewidth=2pt,
  frametitleaboveskip=\dimexpr-\ht\strutbox\relax,
  skipabove=4pt
}
\begin{mdframed}[]\relax%
  \scriptsize\ttfamily 
  }{\end{mdframed}}

\definecolor{rqs}{HTML}{aa5239}
\newenvironment{standoutrqs}[1][]{%
  \mdfsetup{%
    frametitle={%
        \tikz[baseline=(current bounding box.east),outer sep=0pt]
        \node[anchor=east,rectangle,fill=rqs]
        {\strut\footnotesize\textcolor{white}{#1}};}}
  \mdfsetup{innertopmargin=4pt,linecolor=rqs,%
    linewidth=2pt,topline=true,%
    frametitleaboveskip=\dimexpr-\ht\strutbox\relax,
    skipabove=4pt
  }
  \begin{mdframed}[]\relax%
    \label{rq:#1}}{\end{mdframed}}

\definecolor{findings}{HTML}{28784D}
\newenvironment{standoutfindings}[1][]{%
  \mdfsetup{%
    frametitle={%
        \tikz[baseline=(current bounding box.east),outer sep=0pt]
        \node[anchor=east,rectangle,fill=findings]
        {\strut\footnotesize\textcolor{white}{#1}};}}
  \mdfsetup{innertopmargin=4pt,linecolor=findings,%
    linewidth=2pt,topline=true,%
    frametitleaboveskip=\dimexpr-\ht\strutbox\relax,
    skipabove=4pt
  }
  \begin{mdframed}[]\relax%
    \label{finding:#1}}{\end{mdframed}}

\newcommand{\promptboxautorefname}{Prompt}
\definecolor{hlcolora}{HTML}{FFFF66}
\newcommand{\hla}[1]{%
  \colorbox{hlcolora}{\{#1\}}%
}
\definecolor{hlcolor}{HTML}{83C0A0}
\newcommand{\hlb}[1]{%
  \colorbox{hlcolor}{#1}%
}


\newcommand{\totalProjects}{609\xspace}
\newcommand{\totalCommits}{725 \xspace}
\newcommand{\totalMavenCommits}{304\xspace}
\newcommand{\totalAntCommits}{173\xspace}
\newcommand{\totalGradleCommits}{248\xspace}
\newcommand{\totalRefactoringCategories}{24\xspace}
\newcommand{\totalBuildSpecificRefactoringCategories}{8\xspace}
\newcommand{\totalOtherRefactoringCategories}{16\xspace}
\newcommand{\totalTechnicalDebts}{5\xspace}

\newcommand{\totalCategories}{6\xspace}

\newcommand{\toolFoneScore}{0.76\xspace}


\newcommand{\toolName}{BuildRefMiner\xspace}
\newcommand{\rqOneContent}{Which  refactoring types are developers applying to build code?\xspace}
\newcommand{\rqTwoContent}{How are build refactorings linked to technical debt?
\xspace}


\newcommand{\rqThreeContent}{What is the effectiveness of our tool \toolName in identifying build refactorings?\xspace}

\newcommand{\buildfile}{build script\xspace}
\newcommand{\buildfiles}{build scripts\xspace}

\newcommand{\buildrefactoring}{build refactoring}

\maketitle
\begin{abstract}
In modern software engineering, build systems play the crucial role of facilitating the conversion of source code into software artifacts. Recent research has explored high-level causes of build failures, but has largely overlooked the structural properties of build files. Akin to source code, build systems face technical debt challenges that hinder maintenance and optimization. While refactoring is often seen as a key tool for addressing technical debt in source code, there is a significant research gap regarding the specific refactoring changes developers apply to build code and whether these refactorings effectively address technical debt.

In this paper, we address this gap by examining refactorings applied to build scripts in open-source projects, covering the widely used build systems of Gradle, Ant, and Maven. Additionally, we investigate whether these refactorings are used to tackle technical debts in build systems. Our analysis was conducted on \totalCommits examined build-file-related commits.
We identified \totalRefactoringCategories build-related refactorings, which we divided into \totalCategories main categories. These refactorings are organized into the first empirically derived taxonomy of build system refactorings. Furthermore, we investigate how developers employ these refactoring types to address technical debts via a manual commit-analysis and a developer survey. In this context, we identified \totalTechnicalDebts technical debts addressed by these refactorings and discussed their correlation with the different refactorings. Finally, we introduce BuildRefMiner, an LLM-powered tool leveraging GPT-4o to automate the detection of refactorings within build systems. We evaluated its performance and found that it achieves an F1 score of \toolFoneScore across all build systems.

This study will serve as a foundational building block for guiding future research and practice in the maintenance and optimization of build systems. BuildRefMiner and the replication package for this study are available at \cite{material}
\end{abstract}

\section{Introduction}
\label{sec:Introduction}

    
Build systems are responsible for transforming source code into executable programs by coordinating the execution of various tools, ranging from compilers to code analyzers~\cite{Hassan2017,mcintosh2010msr,mcintosh2011icse}. Build systems, such as Maven~\cite{varanasi2019introducing}, Ant~\cite{matzke2003ant} and Gradle~\cite{davis2019gradle} are commonly employed in the development of large software projects to automate the process of compiling, packaging, and testing software products. However, with this wide range of capabilities and flexibility, a lot of complexity can emerge in build code.
Indeed, prior research~\cite{mcintosh2012evolution, kerzazi2014automated, seo2014programmers, kumfert2002software} has demonstrated how configuring build systems can frequently lead to challenges in their maintenance and result in delays in software development projects.
Seo et al.~\cite{seo2014programmers} demonstrated that up to 37\% of builds conducted at Google experience failure, while Kumfert et al.~\cite{kumfert2002software} further estimate that build maintenance imposes a 12\% overhead on the development process, distracting developers from their main tasks.
This highlights the importance of a robust, clean, and well-maintained build system to facilitate seamless development tasks, thereby preventing them from becoming arduous and time-consuming, consequently affecting the overall efficiency of the software.

Refactoring, defined as the restructuring of existing code to improve its quality without altering its outward behavior~\cite{fowler2018refactoring}, is frequently proposed as a method to achieve this objective.
However, Refactoring build systems, despite their importance in software development, remains an area with limited understanding~\cite{nejati2023code,ivers2024mind,aljedaani2024boring}. To the best of our knowledge, no empirically validated taxonomy of build refactorings currently exists. Moreover, no analysis has been conducted to connect build refactorings to the technical debts they may mitigate. Technical debt (TD) is a metaphor that describes the lower-quality code, which represents a trade-off between the short-term benefits of rapid delivery and the long-term value of software~\cite{avgeriou2016managing}.

Currently, practitioners have limited access to specific guidance on how to apply build-related refactorings and organize their build code.
Knowing the types of refactorings and TDs that may occur within build systems can shape developer guidance when it comes to maintaining their existing build files and can support the development of tools that can automatically detect and recommend refactoring opportunities.
To highlight the relevance of refactorings in build files, we provide the example in~\autoref{lst:DRY}, which demonstrates a Don't Repeat Yourself (DRY) refactoring applied to a Gradle file, to address Code Duplication TD. 
Here, 'Connect' and 'Insert' tasks are streamlined using a loop, which reduces repetitive task definitions,
thus minimizing redundancy by avoiding repetitive setups for each task, making the code more maintainable.
\vspace{-0.6cm}

\begin{center}
\begin{varwidth}{0.5\textwidth}
\vspace{-0.25cm}
\begin{lstlisting}[language=diff,caption={Commit \textit{\textsl{560850d}} in \textit{\textsl{biginsight-examples}} Project: An Example of \textbf{DRY} Refactoring Type in Gradle.},label={lst:DRY},numbers=left,frame=lines, escapechar=\!]
!\GreenL!['Connect', 'Insert'].each { taskName -> 
!\GreenL!task "$taskName" (type: JavaExec) {environment 'username',..}
!\RedL!task('Connect', type: JavaExec) {environment 'username', ..}
!\RedL!task('Insert', type: JavaExec) {environment 'username', ..}
\end{lstlisting}
\end{varwidth}
\end{center}





In this paper, we address the knowledge gap on build refactoring types by conducting an empirical study on build refactorings in open-source projects. Our analysis includes building a taxonomy of build-related refactorings, an investigation into which TDs these refactorings address, and the creation of \toolName to automatically detect build refactorings in past commits. In this context, we address the following research questions:






\textbf{RQ1: \rqOneContent} This RQ aims to build a taxonomy of build refactorings. We were able to develop a taxonomy of \totalRefactoringCategories build refactoring types classified into \totalCategories main categories. \totalBuildSpecificRefactoringCategories refactoring changes are Build-specific such as Dependency Organization and Synchronizing Shared Build Properties refactorings.


%

\textbf{RQ2: \rqTwoContent}This RQ aims to uncover which TDs can motivate developers to implement build refactoring operations.
In total, we were able to extract \totalTechnicalDebts TDs linked to refactoring categories. For example, DRY addresses the TD  Code Duplication \& Redundancy.


\textbf{RQ3: \rqThreeContent} To provide guidance to future research, we developed \toolName,
to automatically detects build refactorings in the commit history of a project. 
We were able to achieve an F-1 score of \toolFoneScore in detecting build refactorings.

The main contributions of this paper are:
\begin{enumerate}

    \item The first dataset on refactorings in Build systems. The dataset can be further utilized in future research on developing tools and techniques for detecting and recommending refactoring opportunities.
    
    \item The first quantitative and qualitative study on refactoring changes in multiple build systems: Maven, Ant, and Gradle. We propose a rich taxonomy of a total of \totalRefactoringCategories refactorings divided into \totalCategories main build refactoring categories. Furthermore, we link 
20 of these categories with TDs they address via 85 commit messages and 60 developer responses. 
    
     

     \item \toolName, which we develop for automatic build-refactoring identification, enabling researchers to detect refactoring patterns within build systems more efficiently.

\end{enumerate}

\vspace{-0.2cm}
\section{Background}
\label{sec:background}

\subsection{Build Systems}

Build automation is a critical part of modern software development\cite{mcintosh2012evolution}. Build tools provide flexible ways to model, manage, and maintain complex software projects. 
They rely on configuration files to define dependencies, specify build tasks, and orchestrate build process workflows \cite{Muschko_2014,Gradle,Ant,Maven,ghammam2023dynamic,ghammam2024efficient}. Within this work, we specifically focus on Ant, Maven, and Gradle, widely-used build tools that provide a good representation of the different approaches to build automation.

\textbf{Ant} uses an XML-based  highly customizable  build script to define tasks and dependencies. 
Ant does not have predefined build lifecycle phases. Instead, build targets can be connected to indicate dependencies.
\textbf{Maven} also uses an XML build script, centered around its project object model (POM.xml), which defines project configuration, dependencies, and settings. 
Rather than customizable tasks, Maven relies on plugins bound to these phases to execute goals in a consistent sequence. 
Finally, \textbf{Gradle} builds on Ant and Maven using a Groovy or Kotlin domain-specific language (DSL) for its build scripts. This allows flexible modeling of build requirements in a declarative, extendable way.
Gradle build files consist of components like plugins, dependencies, configurations, tasks\footnote{units of work or action in a build system (e.g., compiling, packaging).}, and properties. 

While Ant, Gradle, and Maven are commonly associated with Java, these tools support other programming environments, such as Kotlin, PHP, and Android. The inclusion of such projects, detailed in our appendix \cite{material},  ensures our findings are representative of diverse ways of build-systems usage in practice.
\vspace{-0.1cm}
\subsection{Refactoring and Technical Debt}

Technical debt is defined as the cost of additional work created by choosing an easy solution now instead of using a better approach that would take longer \cite{10.1145/2507288.2507326}. It is a metaphor that describes the consequences of poor software design and implementation decisions. Technical debt can manifest in various forms in Build files, such as design flaws, and outdated dependencies. Over time, technical debt can accumulate and slow down development, increase maintenance costs, and reduce software quality \cite{10.1145/2507288.2507326}. Refactoring is defined as the process of restructuring existing code without changing its external behavior to improve readability, maintainability, and extensibility \cite{10.1145/2507288.2507326}. Refactoring can help reduce technical debt by improving the design and structure of the codebase. In the context of build code, refactoring can involve simplifying build scripts, removing duplication, and improving maintainability~\cite{duncan2009thoughtworks}. 


\vspace{-0.1cm}
\section{\textbf{Methodology}}
\label{sec:ApproachOverview}

\begin{figure}[ht!]
    \centering
    \includegraphics[width=0.89\linewidth]{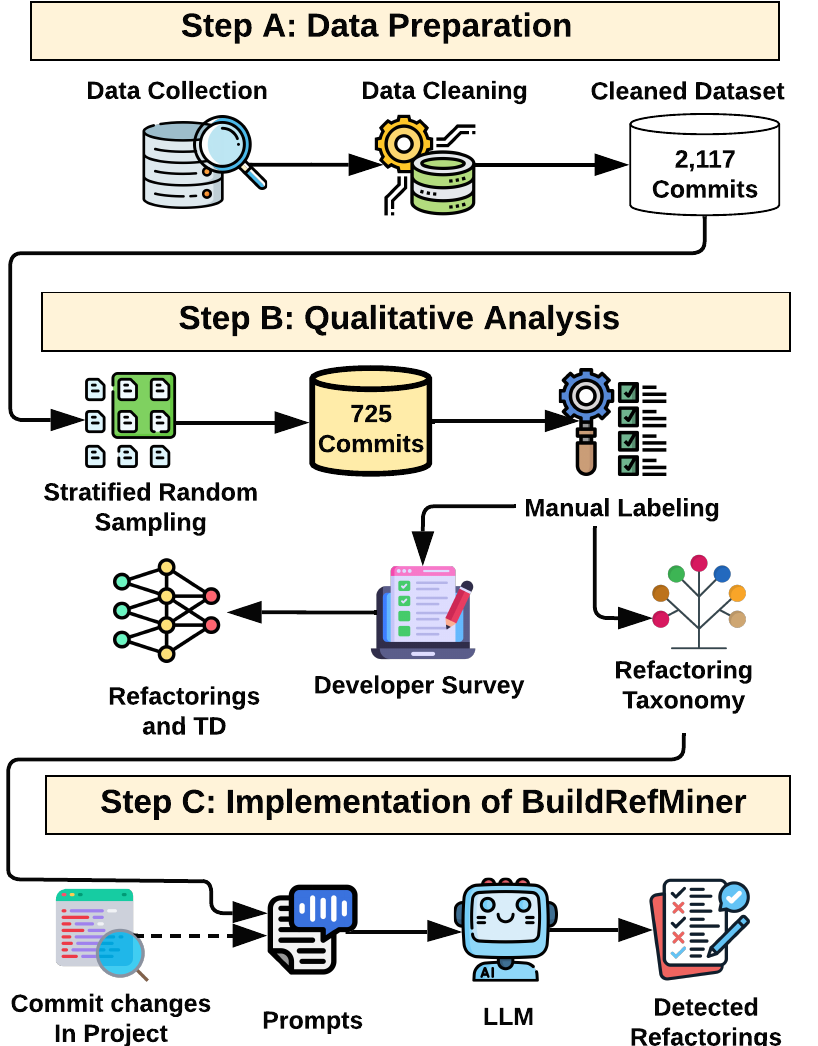}
    \caption{Approach Overview}
    \vspace{-0.5cm}
    \label{fig:ApproachOverview}
\end{figure}

In this section, we describe our research methodology.~\autoref{fig:ApproachOverview} provides an overview of the process, which is composed of three primary phases: Data Preparation, Quantitative Analysis, and Implementation of BuildMiner. 

\subsection{Data Preparation}
\subsubsection{Data Collection}
In this step, our goal is to have a gold set of commit changes in which the developers explicitly report the refactoring activity.

To achieve this, we set out to retrieve commits that involve refactorings and are associated with  Gradle, Maven, and Ant.  We utilized BigQuery~\cite{fernandes2015bigquery}, an extensive collection of open-source GitHub projects, via the SQL query presented in~\autoref{lst:query}.
We utilized a keyword-based mechanism to filter out all entries that do not contain the keyword ‘refactor*’ in the commit message. 
We use * to capture extensions like refactors, refactoring, etc.
The keyword-based approach has been widely used in previous studies related to identifying refactoring changes \cite{alomar2021refactoring,alomar2024behind,alomar2022exploratory}, as it allows us to prune the search space to only consider code changes whose
documentation matches a specific intention.
The choice of ‘refactor*’,
was specifically made to reduce the occurrence of false positives \cite{rosen2016mobile,alomar2024behind,alomar2021we}. Furthermore, we mention variations of file names corresponding to Maven, Ant, and Gradle in our query. 
Overall, these steps
helps ensure that the extracted refactorings are more likely to be build-related. Through this initial filtering, we collected a total of 5k commits. 


\begin{lstlisting}[language=SQL, caption={Extracting Refactoring-related Commits}, label={lst:query},  basicstyle=\ttfamily\footnotesize][ht!]
    SELECT * FROM "bigquery-public-data.github_repos.commits" WHERE (message LIKE '%refactor%') AND (message LIKE '%pom.xml%' 
    OR message LIKE '%build.xml%' OR message LIKE '%build.gradle%');
\end{lstlisting}

\subsubsection{Data Cleaning}


In this step, we omit commits pertaining to repositories derived from other repositories, such as forked repositories, to prevent any redundancies and resulting potential biases  in our study. In addition,
we exclude projects that no longer exist, as they may be de-listed from GitHub but still exist in BigQuery.
This left 
a total of 2453 commits, 691 correspond to Gradle, 1450 correspond to Maven and 312 correspond to Ant. Given the substantial size of the data and the considerable effort and time required for its manual examination, we applied a stratified random selection process to select a diverse and representative sample of build-related commits, following the recommendations of Levin et al.~\cite{levin2019towards}. The stratification criteria was the build systems to which the commits correspond. For each build system, we selected a sample with a confidence level of 95\% and a Margin of Error of 5\%, which was determined to 248 are for Gradle, 173 for Ant, and 304 for Maven, for a total of 725 commits.
These commits originated from 609 projects, which had an average 86K and median 9.7K lines of codes, an average 160 and  median 1 forks, and finally an average 606 and median 5 stars.


These projects used a varied set of 30 programming languages. For example: 487 projects used Java, 51 used Kotlin, 19 used PHP, among other programming languages, as outlined in the appendix~\cite{material}.

\vspace{-0.1cm}
\subsection{Qualitative Analysis methodology}
\label{subsec:Refactoring-Identification}
\vspace{-0.1cm}
\subsubsection{Commit Analysis Methodology}
\label{subsubec:taxonomy_methodology}
Given that this study represents the initial investigation into the development of code-specific refactorings for build systems, it was imperative to conduct a manual examination of commits to identify the different refactorings as well as any relationship they may have with TD. The labeling was based on an open-coding process that allowed refactoring types to emerge from the data rather than preconceptions. We purposefully avoided enforcing existing refactoring operations since some of them may be limited to other artifacts like source code, while others may be specific to build code. We also
followed the recommendations outlined by Usman et al. concerning the construction of taxonomies~\cite{usman2017taxonomies}. 

First, during the planning phase, three co-authors agreed on the area of focus being \buildrefactoring. The main aim as the identification of the different types of refactoring that developers use in \buildfiles and to ascertain whether these were utilized to address technical debt. Finally, the classification framework is represented as a tree. Although a build refactoring taxonomy had been proposed in the literature by Simpson et al. \cite{duncan2009thoughtworks}, it was not utilized in this process to avoid bias, as itlackes empirical validation, and was technology-specific to an old version of Ant. 

  Second,  the identification and extraction phase was performed over two rounds. For the first round (Identification), two of the co-authors with refactoring and \buildfiles experience
separately observed all the build commits diffs in order to identify any refactoring changes and explicit mentions of TD. Their primary criterion was the definition of refactoring as "restructuring existing code to improve its quality without altering its external behavior". If a commit changes behavior, they don’t consider it a refactoring. To systematically identify the observed refactorings, they used the build systems' official documentation, refactoring principles, commit names, and messages.

The two co-authors kept a shared record of high-level descriptions of the refactoring changes serving as a dynamic reference, containing for each refactoring candidate a textual definition, a candidate-name, and an illustrative labeler-written code stub (to reduce bias/information leakage between the labelers
in order to avoid any redundant definitions).

For the second round (categorization), the third co-author with previous \buildfiles experience was asked to rejoin this process. This was to minimize the bias, as this author did not participate in \textcolor{black}{the identification of refactoring changes} in the previous round. These three co-authors then categorized refactoring operations identified in the first round and mentions of TD \textcolor{black}{(a refactoring commit is only linked to a TD category if it is explicitly mentioned in commit messages or developer feedback)}. These rounds were performed over one month to avoid labeling fatigue.

After the two labeling rounds concluded, Fleiss' Kappa score \textcolor{black}{reflecting the agreement on both the identification of refactoring commits and their classification into specific types,} was calculated at 0.76
signaling high agreement~\cite{Campbell_Quincy_Osserman_Pedersen_2013}. Then, three rounds of consensus meetings were carried out to resolve any disagreements.


Third, for the design and construction phase, a card-sorting process~\cite{fincher2005making,spencer2004card} was performed by the three authors.
This process grouped refactoring changes of similar characteristics into \totalRefactoringCategories categories. 
Subsequently, they refined this classification by determining \totalCategories high-level main categories, for an easier generalizability and usability, giving it the format of a tree. Furthermore, we  provide  specific examples along with the definitions of the categories in order to avoid definition bias.
This taxonomy is presented and detailed in~\autoref{subsec:RQ1}.

Finally, for the Validation phase, a fourth co-author who possesses extensive expertise in the field of build systems confirmed the applicability of the various categories, ensuring they reflect significant recurring patterns and constitute valuable knowledge for developers.


\textcolor{black}{Regarding the relationship between refactorings and TD, Peruma et al. \cite{peruma2022refactoring} discuss how refactoring does not always address TDs. Indeed, 89\% of refactoring commits they analyzed did not remove TD.} Therefore, the authors found that this process alone was insufficient to establish a link between build refactorings and technical debt. We address this in~\autoref{subsec:survey}.

\subsubsection{Developer Opinion Collection}
\label{subsec:survey}
Concerning the uncovering of links between the identified refactoring categories and TDs, a main challenge was the lack of documentation,
along with ambiguous descriptions, which made it difficult to identify the prevalence of TD in the context of \buildrefactoring. Hence, we opted for a conservative approach, where we only considered TD as being addressed by the refactorings if it was explicitly mentioned in the commit message or code comments. However, this approach may have led to the underestimation of the actual number of instances of TD repayment via \buildrefactoring .  To compensate for this shortcoming, and in line with previous studies \cite{tang2021empirical,alomar2024behind,baltes2016worse,nauman2017survey,fowler2013survey}, we performed a cold-calling-based survey. We emailed and send direct-messages to the commit authors to enquire about the different refactoring instances and whether they were linked to a TD. This survey was open-ended, and only took the form of one question: The motivation behind the applied refactoring changes. This was done to minimize the burden on the developers. This survey was conducted over a period of one month.

In total, we identified 85 commit messages/descriptions that clearly describe the motivation behind applying the refactoring changes. Furthermore, we received survey responses from 60 developers, out of 250 originally contacted, achieving a 24\% response rate, in line with the average response rate of 15-30\%, for software engineering surveys \cite{kitchenham1995case,pacheco2008stakeholder,ciolkowski2003practical}. The survey responses provided us with additional insights into the refactoring changes and their relationship with technical-debt. For example, commit~\cite{ref0} contains an \textit{Extract Module} refactoring , which was accompanied with the ambiguous commit message '\textit{build.gradle refactoring}'. 
The developer clarified that this
refactoring was used to address the modularity, and provided us with the following explanation: \textit{build.gradle was too long (500 lines) and divided the functions...for better modularity}—thus clarifying the link  between the different refactorings in this commit and TD. The results of this process and this survey are detailed in~\autoref{subsec:RQ2}.

\vspace{-0.3cm}
\subsection{Implementation of \toolName}
\label{subsec:Refactoring-Tool-Builidng}
\vspace{-0.1cm}
As mentioned in~\autoref{subsec:Refactoring-Identification}, the difficulty and time-consumption of
manually identifying instances of build refactoring highlights the need for automated, build-specific refactoring discovery. 
Hence, we design \toolName, a tool that can automatically analyze commits that affect build files to extract any refactoring operations that may have occurred. Currently, it utilizes GPT-4o, but it can be configured to use other LLMs. It utilizes LLMs due to their adaptability and proclivity with source-code analysis \cite{nam2024using,bairi2024codeplan,li2024extracting,patil2024review}. We utilized Prompt-Engineering practices \cite{giray2023prompt,sahoo2024systematic} while building this tool, specifically Zero-shot and One-Shot prompting. 

The prompt, a snippet of which is shown in~\autoref{fig:system-prompt}, is composed of a system instruction, a list of the names and definitions of the build refactorings we discovered, inserted at the location of the text highlighted in green. As part of the One-Shot variant of the prompt, a code snippet example demonstrating each refactoring type is given along with each definition. Then, we introduce the new commit changes for analysis under the commit variable highlighted in yellow. \textcolor{black}{This commit placeholder is composed of the full diff of that commit, including the name and path of the build file(s) impacted (e.g., src/pom.xml), Modified Lines: additions and deletions as commonly seen in Version-Control-Systems (e.g., git).}
We developed three distinct one-shot learning prompts, each tailored with examples specific to the build systems Gradle, Maven, and Ant for each refactoring type, where the build system and delimiter variables are changed depending on the build system being analyzed.

\begin{figure}
\begin{promptbox}{Prompt}
    Task: Given the commit changes below that are applied on \hla{build system} build scripts which will be delimited with \hla{delimiter} characters, identify any occurrences of the listed refactoring types.

    Provide the results strictly in JSON format with the following keys: RefactoringType and Details. 

    If there are multiple refactorings, return them as a list of JSON objects, where each object contains the following:
    - "RefactoringType": The type of refactoring detected.
    - "Details": A Description with further information about the change.

    Here are two examples of the format I expect:
    ....

    If no refactorings are detected, return the message: "No refactorings found."

    This is a list of 24 refactoring types in build files:

  \hlb{...}
  
  \hla{commit}

\end{promptbox}
  \vspace{-0.3cm}
  \caption{\toolName Prompt}
  \label{fig:system-prompt}
  \vspace{-0.4cm}
\end{figure}

We evaluated the performance of both prompting approaches, and we detail the results of this evaluation in \autoref{subsec:RQ3}. While it is possible to use static analysis techniques to implement this tool, the reliance on LLMs has allowed us to quickly 
utilize, evaluate, and improve \toolName. Furthermore, this implementation is more flexible and easier to extend for  other build systems, unlike a static-analysis tool that would need to be customized to support every specific build tool. We provide \toolName as an accompaniment to our taxonomy and as a proof of concept concerning the relevance and importance of build refactorings and to facilitate future work. The implementation of \toolName with the usage of static analysis and its comparison with the LLM implementation can be the subject of an interesting future work.

\vspace{-0.1cm}
\section{Study Results}
\label{sec:evaluation}
\subsection{RQ1:\rqOneContent}
\label{subsec:RQ1}
\begin{figure}[h]
    \centering
    \includegraphics[width=\linewidth]{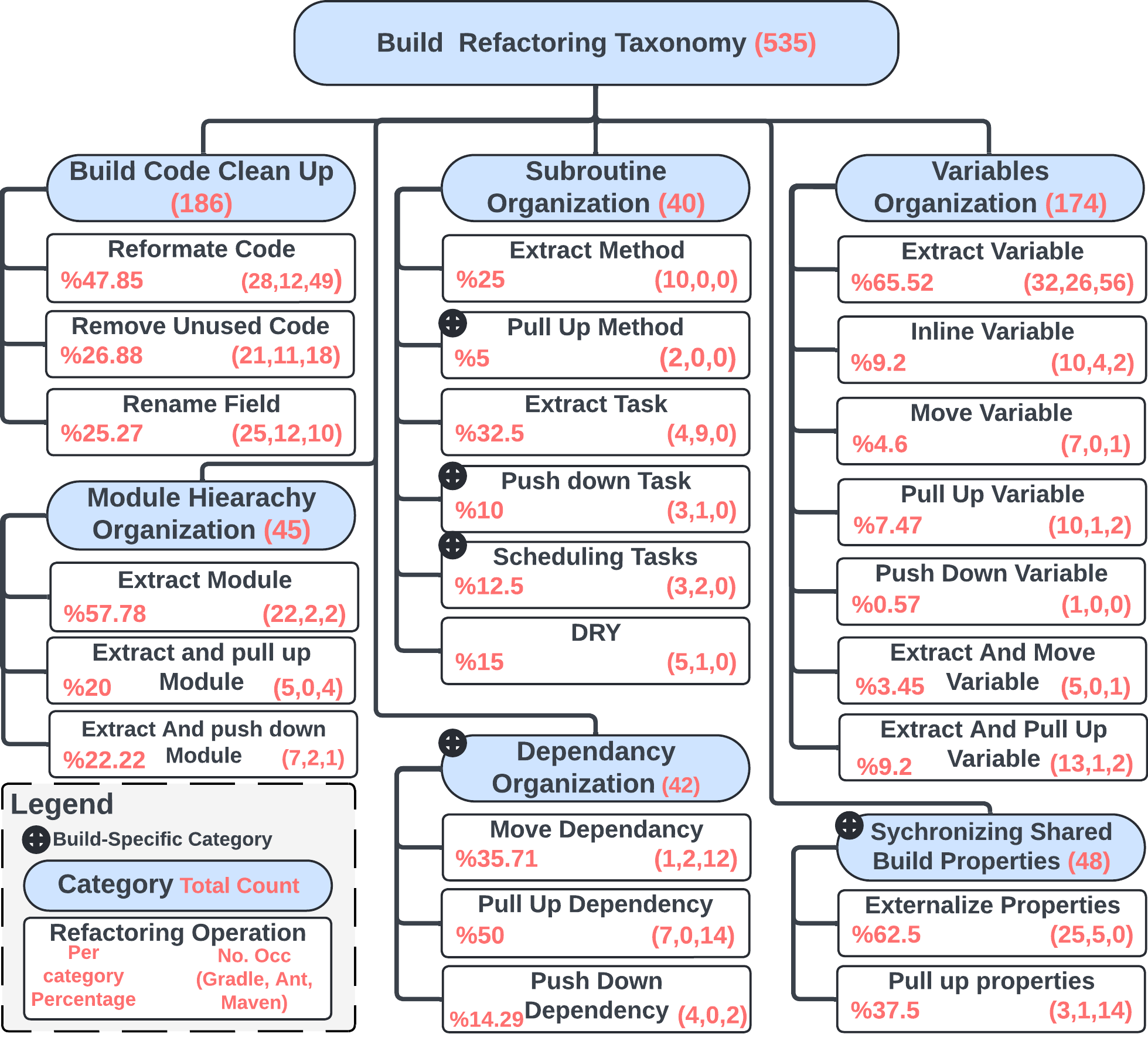}
    \caption{Build-related refactorings taxonomy}
    \label{fig:Taxonomy}
    \vspace{-0.4cm}
\end{figure}

\vspace{-0.1cm}
To uncover and categorize the Build refactoring types that are present within \buildfiles, we followed the methodology discussed in \autoref{subsec:Refactoring-Identification} and carried out a manual analysis of 725 commits. We discovered that 32\% of them were false positives, as they did not contain any discernible refactorings. The primary cause of the inclusion of these false positives was the use 
which sometimes involved the addition or modification of existing functionality~\cite{kim2014empirical, bauer2014make}.
This left a total of 403 true-positive Build-refactoring-related commits.
\autoref{fig:Taxonomy} represents the taxonomy that we have generated based on a parent-child hierarchy.
These refactoring types have been organized into \totalCategories main categories for ease of classification and analysis. The tuple under each refactoring type indicates the number of refactoring changes we discovered from Gradle, Ant, and Maven respectively. Each percentage value 
represents the proportion of refactorings within a specific category out of the total of identified refactorings of the same main category. 
The classification into \totalCategories main categories was based on the scope of their impact on the build code. These ranged from broad, project-level impacts (`Code Clean Up') to more specific levels, including Module (`Module Hierarchy Organization'), Method/Task (`Subroutine Organization'), Dependencies (`Dependency Organization'), Shared Properties (`Synchronizing Shared Build Properties'), and Local Variables (`Variables Organization'). 
The different subcategories offer greater specificity as they represent the different kinds of refactoring identified.
In total, we were able to develop a taxonomy of \totalRefactoringCategories build refactoring types. 
%
In the rest of this section, we discuss each of the Build-related Refactoring categories as well as their sub-categories by giving detailed definitions. In addition, we provide code examples to mitigate the potential bias inherent in definitions. It is notable that these definitions may be extended for application in other code artifacts.

\subsubsection{\textbf{Build Code Clean Up}} This particular category encompasses alterations that serve to improve the cleanliness, readability, and comprehensibility of the build script. The detected refactoring operations represented the most prevalent refactorings for the three build systems, accounting for 34.77\% of all refactorings. Three subcategories were observed:

\textbf{1.1 Reformat Code (RC)} This refactoring makes up 47.8\% (89/186) of the Build Code Clean Up category. It involves aligning code with specific style conventions, adjusting indentation, reorganizing code blocks, and ensuring consistent coding style for clarity. Listing 3 shows an example in a Gradle script, switching to the plugins DSL for readability and modern best practices.



\textbf{1.2 Remove Unused Code (RUC)} This removes redundant or unused build code, accounting for 26.88\% (50/186) of the Build Code Clean Up category.

\textbf{1.3 Rename Field (RF)}: This refactoring, 25.27\% (47/186) of the total, involves renaming build components (e.g., files, tasks, methods) to improve code clarity.



\subsubsection{\textbf{Module Hierarchy Organization}} This particular category pertains to the process of dividing the build script into separate modules (build files), with each build file being assigned a specific role or functionality. These modules are then arranged hierarchically, taking into consideration their dependencies and relationships. The refactoring operations that were identified collectively accounted for 8.41\% of the total refactorings. The study identified three subcategories:

\begin{center}
\begin{varwidth}{0.5\textwidth}
\begin{lstlisting}[language=diff,caption={Commit \textit{\textsl{f833d19}} in \textit{\textsl{android-gradle-jenkins-plugin}} Project: An example of Reformat Code Refactoring Type in Gradle.},label={lst:reformat},numbers=left,frame=lines, escapechar=\!]
!\GreenL! plugins {
!\GreenL!	id 'jacoco'}
!\GreenL!	id 'groovy'}
!\RedL! apply plugin: 'jacoco'
!\RedL! apply plugin: 'groovy'
\end{lstlisting}
\end{varwidth}
\end{center}

\vspace{-0.2cm}

\textbf{2.1 Extract Module (EM)} represents 57.78\% (26/45) of the refactorings in this category, which involves extracting and relocating functionality or responsibilities into a new build file within the same hierarchy. Notably, this refactoring type is more common in Gradle (22 occurrences) than in Maven and Ant (2 occurrences each), making it one of the most prevalent refactorings found in Gradle. Commit~\cite{ref1} illustrates an example of Extract Module. In this commit, developers extracted a segment of the parent 'build.gradle' script detailing multiple functionalities and relocated it to a new build file within the same hierarchy, 'publish.gradle'. The initial build file utilizes apply from 'publish.gradle' to invoke the extracted responsibilities.

\textbf{2.2 Extract And Pull Up Module (EPUM)} represents 20\% (9/45) of the refactorings, it entails moving functionality or responsibilities from a more specialized module or class (child) to a newly created more generalized or parent module. Commit~\cite{ref2} illustrates an example of Extract and Pull Up Module in Maven. This change extracts a segment of a build code that delineates Maven properties from "lib/pom.xml and relocate them to a superior newly created Maven file "pom.xml".

\textbf{2.3 Extract And Push Down Module (EPDM)} represents 22.22\% (10/45) of the refactorings, it shifts functionality or responsibilities from a general or parent module to a more specific one newly created (child).


\subsubsection{\textbf{Subroutine Organization}}

This category covers operations modifying subroutines, specifically tasks or methods, primarily in Gradle and, to a lesser extent, Ant. It is noteable that  Maven lacks comparable concepts. In Gradle, tasks represent actions, and methods configure them. Ant has predefined XML tasks but no methods. Therefore, refactoring related to methods applies solely to Gradle, while task-related refactoring includes both Ant and Gradle but excludes Maven. This category accounts for 7.48\% of all identified refactorings.
In total, we discovered two refactorings that focus on methods, and four refactoring types that are focused on tasks.

\textbf{3.1 Extract Method (EM)} This type represents 25\% (10/40) of the Subroutine Organization category. It entails the identification of a cohesive code fragment and its subsequent relocation into a newly created method or function. Subsequently, the original code is substituted with an invocation of the recently generated method.
Commit~\cite{ref3} provides an example of the Extract Method refactoring applied to a Gradle script. In this instance, the methods \textit{createChecksum()} and \textit{createExecutable()} were extracted from the \textit{makeExecutables} task and invoked within the same task (Lines 34-41), improving code extensibility and reusability.

\textbf{3.2 Pull Up Method (PUM)} accounts for 5\% (2/40) of the total refactorings in this category. It involves moving an existing method to the relevant Parent build file. An example of this refactoring is illustrated in commit~\cite{ref4} where \textit{uploadArchives} method has been pulled up from a subfile \textit{concourse-cli/build.gradle} to the parent file \textit{build.gradle}.



\textbf{3.3 Extract Task (ET)} accounts for 32.5\% (13/40) of the total refactorings in this category. It involves identifying a cohesive code fragment, relocating it into a new task, and replacing the original code with a call to the newly created task. 
\autoref{lst:extracttask} is an example of Extract Task refactoring in Gradle. It defines a new task, resolveDependencies, which is set as a prerequisite for the publish task.
Commit~\cite{ref5} serves as an illustration of the Extract Task refactoring type within the Ant build system. In this instance, eleven defined tasks (such as \textit{start}, \textit{start-secure}, and \textit{start-batch}) exhibited identical behaviors, leading to code duplication. To eliminate this redundancy, a new task, \textit{exec-scipio-jar}, was introduced (line 1033) to consolidate and encapsulate the shared behavior of these tasks.

\begin{center}
\vspace{-0.2cm}
\begin{varwidth}{0.5\textwidth}
\begin{lstlisting}[language=diff,caption={Commit \textit{\textsl{4f03800}} in \textit{\textsl{droidmate}} Project: An example of Extract Task Refactoring Type in Gradle.},label={lst:extracttask},numbers=left,frame=lines, escapechar=\!]
!\RedL! publish.dependsOn {
!\GreenL! task resolveDependencies{doFirst {
    project.publishing {....}}}
!\GreenL! publish.dependsOn(resolveDependencies)
\end{lstlisting}
\end{varwidth}
\vspace{-0.2cm}
\end{center}

\textbf{3.4 Push Down Task (PDT)} represents 10\% (4/40) of the refactorings. It entails moving a task defined in the root project's build file to the build file of one or more sub-projects.

\textbf{3.5 Scheduling Tasks (ST)}: This refactoring type represents 12.5\% (5/40) of the total refactorings in this category. It ensures a specific order of execution of tasks, or the execution multiple tasks simultaneously.
In Ant, sorting tasks can be done using \textit{depends} attribute that specifies the order of execution by indicating which tasks need to be completed before the current task begins. \autoref{lst:SchedulingTask2} shows an example of a Scheduling Tasks refactoring in Ant. The changes were in place to ensure that \textit{init} task was executed only after the \textit{clean} task (Line 2).

\textbf{3.6 Don't Repeat Yourself (DRY)}:

This refactoring type is related to the process of consolidating the common behavior of multiple tasks into a single location method to void redundancy. \autoref{lst:DRY} shows an example of the \textit{DRY} refactoring type in Gradle. Initially, tasks such as Connect and Insert were explicitly configured, each possessing its own set of parameters such as environment variables (Line 3-4). In the revised iteration, the identical configurations are dynamically applied to tasks by iterating through a list that includes the task names Connect and Insert (Line 1). As a consequence, the script becomes more succinct by eliminating repetitive configurations and replacing them with a generalized loop. 

\begin{center}
\vspace{-0.4cm}
\begin{varwidth}{0.5\textwidth}
\begin{lstlisting}[language=diff,caption={Commit \textit{\textsl{717417f}} in \textit{\textsl{aDTN-platform}} Project: An example of Scheduling Tasks Refactoring Type in Ant.},label={lst:SchedulingTask2},numbers=left,frame=lines, escapechar=\!]
!\RedL!<target name="init"
!\GreenL!<target name="init" depends="clean">
    <mkdir dir="${bin}"/>.....
</target>
\end{lstlisting}
\end{varwidth}
\vspace{-0.1cm}
\end{center}


\subsubsection{\textbf{Dependency Organization}} 

This particular category is a build-specific category, concerned with the reorganization of dependencies between build files, accounting for 7.85\% of the overall count of identified refactoring changes. It plays a crucial role in determining the order and manner in which build dependencies are built and integrated. 
The study conducted identified three distinct types:

\textbf{4.1 Move Dependency (MD)} It refers to the transfer of dependency between two build files within the same hierarchy. It accounts for 35.71\% (15/42) of the overall count of identified refactoring changes in this category, with 12 occurrences for Maven, compared to 1 and 2 occurrences each for Gradle and Ant respectively.

\textbf{4.2 Pull Up Dependency (PUD)} It involves moving a dependency from a sub-build file to its parent Build file; It accounts for 50\% (21/42) of the overall count of identified refactoring changes in this category. Notably, this refactoring is more frequent in Maven, with 14 occurrences, compared to 7 and 0 occurrences each for Gradle and Ant respectively. Commit~\cite{ref7} exemplifies this type of refactoring by consolidating various dependencies, including the Spring Boot Gradle plugin and the Kotlin Gradle plugin, that were previously declared across multiple sub-build files. These dependencies were eliminated from the individual sub-build files and migrated to the parent build file \textit{build.gradle}, specifically within lines 1-38, to enhance maintainability and reduce redundancy.


\textbf{4.3 Push Down Dependency (PDD)} It accounts for 14.29\% (6/42) of the overall count of identified refactoring changes in this category. It entails moving a dependency from a parent file to a child or specific build file.

\subsubsection{\textbf{Synchronizing Shared Build Properties}} This category is another discovered Build-specific refactoring category and represents 8.97\% of the overall count of the identified refactoring types. It refers to the process of ensuring that multiple parts or modules of a build system use a consistent set of properties. These properties can be configurations, versions, paths, or any piece of data that affects how the build process operates. When projects grow in complexity and include multiple components or modules, it's essential to maintain a single source of truth for shared properties to avoid inconsistencies.
Two distinct types were observed during the study:

\textbf{5.1 Externalize Properties (EP)}: Accounts 37.5\% (18/48). This type of refactoring focuses on the centralized and harmonized handling of build configurations by extracting environment-specific configurations, credentials, and settings from the build script and application code. This is typically achieved by utilizing external properties or separate configuration files, such as \textit{.properties} files. This type of refactoring is particularly common in Gradle, with 25 instances, compared to 5 and 0 instances for Ant and Maven, respectively.
\autoref{lst:ExternalizeProp1} shows an example of Externalize Properties.
The initial implementation involved hardcoding the version value directly into the script. The revised modifications improve version management efficiency by introducing an external \textit{build.properties} file (Line 2) that encapsulates essential configurations. This approach decouples version management from the build script, allowing users to update configurations easily by modifying the properties file rather than directly altering the script itself. 
Decoupling the components of the build system not only improves the ease of reading, but also enhances the ability to maintain it. This allows developers to make changes to configurations without needing to delve into the fundamental build logic.

\begin{center}
\vspace{-0.4cm}
\begin{varwidth}{0.5\textwidth}
    \begin{lstlisting}[language=diff,caption={Commit \textsl{611c4b3} in \textit{\textsl{YetAnotherBackupMod}} Project: An Example of Externalize Properties Refactoring in Gradle.},label={lst:ExternalizeProp1},numbers=left,frame=lines, escapechar=\!]
!\RedL!version = "1.7.10-0.1"
!\GreenL!ext.configFile = file "build.properties"
!\GreenL!configFile.withReader {def prop = new Properties()
!\GreenL!  project.ext.config = new ConfigSlurper().parse prop}
!\GreenL!version = config.version
\end{lstlisting} 
\end{varwidth}
\vspace{-0.2cm}
\end{center}

\textbf{5.2 Pull Up  Properties (PUP)} Accounts 62.5\% (30/48): pertains to the process of consolidating frequently used configurations and properties within a build file into a centralized location or method within the root build file. This refactoring type is most common in Gradle Maven with 25 occurrences compared to 3 and 1 occurrences for Gradle and Ant respectively. A specimen of this refactoring type is illustrated in Commit~\cite{ref8}. The changes involve migrating configuration elements from a child Maven build file \textit{core/pom.xml} (Line 24-29) to the root \textit{pom.xml} (Line 56-60). By centralizing the source control management metadata in the root file, the project benefits from improved coherence and ease of updates, as any changes to these properties need to be made in only one location.



\subsubsection{\textbf{Variables Organization}} This category constituted the second largest refactoring category after the Code Clean Up  Category, accounting for 32.52\% of the total count of identified refactoring types. It is primarily associated with the organization of variables within a single build file or across the hierarchy of build files. All the identified subcategories of variable refactoring are more frequent in Gradle compared to Ant and Maven. They are as follows: \textbf{Extract Variable(EV)} accounting for (65.52\%), \textbf{Inline Variable (IV)} (9.2\%), \textbf{ Move Variable (MV)} (4.6\%), \textbf{Pull Up Variable (PUV)} (7.47\%), \textbf{Push Down Variable (PDV)} (0.57\%), \textbf{Extract And Move Variable (EMV)} (3.45\%), and \textbf{Extract And Pull Up Variable (EPUV)} (9.20\%).

\textcolor{black}{Out of these refactoring categories, some are that are analogous to other artifacts and adapted to the context of build systems (e.g Pull Up Properties) and some that emerge due to the peculiar nature of build artifacts} (Build-specific) that do not have equivalents in other artifacts
(e.g., source code).
Among the \totalRefactoringCategories refactorings, we identified 8 as Build-specific types. These include the Tasks-related refactorings, under the \textit{Subroutine Organization} main category (Scheduling Tasks, Push Down Task, and Extract Task), and all the refactoring belonging to the
\textit{Dependency Organization}
and \textit{Synchronizing Shared Build Properties} main categories.
We observed that developers  have to specifically adapt to the context of Build files  when applying these refactoring changes. This distinction arises from the nature of their primary concerns, which revolve around the foundational elements of build systems: tasks, dependencies,
and properties.

Comparing our Taxonomy with Simpson et al.~\cite{duncan2009thoughtworks}, the categories we defined are much more technology agnostic, in comparison to some of their technology-specific categories, such as \textit{Move element to Antlib} and \textit{Replace Exec with Apply}. Furthermore, Simpsons' taxonomy does not include some categories we discovered, such as DRY and Code Cleanup. While some categories are indeed similar between the two taxonomies, such as Simpson's Introduce Property File and our Externalize Properties Categories, our taxonomy is organized into empirically validated \totalCategories Main categories that are dependent on scope, and we use naming conventions that are more inline with Refactoring in other domains, which makes our taxonomy easier to understand and utilize.  



\begin{standoutfindings}[Finding 1]
    We created a build refactoring taxonomy composed of \totalRefactoringCategories refactorings types, that fall under \totalCategories main refactoring categories. \totalBuildSpecificRefactoringCategories of the 24 refactorings are specific to build systems.
\end{standoutfindings}

\subsection{RQ2: \rqTwoContent}
\label{subsec:RQ2}
Our analysis of  85 commit messages and  60 developer responses shows that 20 out of the 24 identified refactoring types are associated with technical debt (TD) reduction, accounting for 94.01\% of the total refactoring instances. However, five refactoring types—Move Dependency, Extract and Move Variable, Move Variable, Push Down Task, and Push Down Variable—were not classified as related to TD repayment.
The rationale behind the application of these five refactoring types remains unclear, suggesting the need for further investigation into their role in TD repayment.

\subsubsection{Technical Debt Categories }
\label{subsubsec:RQ2.1}

We extracted \totalTechnicalDebts technical debt categories related to \textit{Extensibility \& Maintainability} , \textit{Modularity}, \textit{Clarity \& Readability},\textit{ Code Duplication \& Redundancy}, and \textit{Security}. We organized the different refactoring types based on their rationale, e.g., technical debts they address as shown in \autoref{tab:technical-debts}. The percentage associated with each technical debt represents the cumulative number of refactorings undertaken to address that particular TD out of the total number of refactorings identified.

\textbf{Clarity \& Readability} is a TD that receives considerable attention during the process of refactoring build files by 39.63\% of the total refactorings. It was linked with the following refactoring types: \textit{Inline Variable}, \textit{Rename Field}, \textit{Reformat Code} and \textit{Remove Unused Code}.
These refactorings were used when code is not clear, and hard to read requiring huge efforts to understand it. For example, Rename field was used to avoid using irrelevant names for build artifacts and to ensure consistency in naming conventions, like  the example commit message 
\textit{"In order for this to become more comprehensive ... the package name needed to be changed"} from the commit~\cite{cm0}.

Reformat Code and Inline Variable were used to ensure 
better readability and understandability as shown in the developer response in~\autoref{fig:dc_reformat} to commit~\cite{dm0}.


\begin{figure}
    \vspace{-0.1cm}
    \begin{devbox}{Developer Response}
        The reason was rather to unify code stale and improve readability.
    \end{devbox}
    \vspace{-0.3cm}
    \caption{Developer Response on Commit~\cite{dm0}}
    \label{fig:dc_reformat}    
    \vspace{-0.1cm}
\end{figure}

\textbf{Extensibility \& Maintainability} was among the most addressed TDs in
build files, and was mainly associated with: \textit{Extract Variable}, \textit{Extract And Move Variable}, \textit{Extract Method}, \textit{Extract Task}, and \textit{Move Task}, repaid by 30.66\% of the total refactorings. 



As software grows and evolves, adding new components or improving existing ones often requires modifying or duplicating elements in the build system, which can lead to challenges in both extensibility and maintainability. Refactoring techniques like Extract Variable, Extract Task, and Extract Method aim to increase the system's abstraction levels and reusability, making future modifications easier and reducing the risk of widespread bugs. This is illustrated by a example commit message \textit{"Task 'retroweaver' moved to the prepare target so it can be reused in several places"} for the commit~\cite{cm1}.

Also, externalizing properties ensures that changes can be made with minimal disruption, enhancing both the flexibility to adapt the system and the ease with which future changes and maintainability efforts as in the example of a developer response shown in~\autoref{fig:cm_consistent}


\begin{figure}
    \vspace{-0.3cm}
    \begin{devbox}{Developer Response}
        Moving properties on its own file gives us the ability to force a dependency(s) version to be consistent
        ...
    \end{devbox}
    \vspace{-0.3cm}
    \caption{Developer Response of Commit~\cite{dm1}}
    \vspace{-0.3cm}
    \label{fig:cm_consistent}    
\end{figure}  

\textbf{Modularity} 
This technical debt is addressed by 14.95\% of the total refactorings, such as 
Extract Module, Extract and pull up Module, Extract and Push Down Module, Push Down Task, Pull Up Method, Pull Up Variable, Extract and Pull Up variable and Push Down Dependency. These refactoring techniques break down complex build files into smaller, modular units, ensuring that each component is situated at the most appropriate level of the build system hierarchy, making management easier, clearer, and reducing errors during updates. This is illustrated by 
commit messages~\cite{ref0}: \textit{"refactor build.xml for centralized usage for all plugins."}, and developers answers such as the ones shown in~\autoref{fig:dm_modular} and \autoref{fig:dm_decouple}.
Extract Module, Extract and pull up Module, Extract and Push Down Module, Push Down Task, Pull Up Method, Pull Up Variable, Extract and Pull Up variable and Push Down Dependency. These refactoring techniques break down complex build files into smaller, modular units, ensuring that each component is situated at the most appropriate level of the build system hierarchy, promoting simplicity, clarity, making it easier to manage, and less complex, and reducing the likelihood of errors when components are changed. This is illustrated by developers answers shown in~\autoref{fig:dm_modular} and \autoref{fig:dm_decouple}, and the commit message \textit{"refactor build.xml for centralized usage for all plugins."} for commit~\cite{ref0}



\begin{figure}
    \vspace{-0.1cm}
    \begin{devbox}{Developer Response}
        ...
        I have decided that build.gradle was too long (500 lines) and divided the functions in it according to responsibilities, for better modularity
        ...
    \end{devbox}
    \vspace{-0.3cm}
    \caption{Developer Response for Commit~\cite{ref0}}
    \label{fig:dm_modular}  
    \vspace{-0.2cm}  
\end{figure}  

\begin{figure}
    \vspace{-0.2cm}
    \begin{devbox}{Developer Response}
        Split dependencies in optional modules (at this time gumetree was growing to much and we started to make it more decoupled)
    \end{devbox}
    \vspace{-0.3cm}
    \caption{Developer Response for Commit~\cite{dm2}}
    \vspace{-0.3cm}
    \label{fig:dm_decouple}    
\end{figure}

\textbf{Code Duplication} Duplicated code and redundant elements make a system more difficult to maintain, prone to errors, and less efficient. This is addressed by 4.20\% of the total refactorings Refactoring techniques such as Pull Up Properties, Pull Up Method, Pull Up Dependency, Extract and Pull Up Variable, and Pull Up Variable focusing on consolidating shared code fragments across various classes into higher-level modules. By pulling up shared elements, these techniques eliminate duplication, making the code more manageable, and reducing the likelihood of inconsistencies. This is illustrated by a commit message \textit{"Reduced duplicate POM XML content"} for commit~\cite{ref9},
where the refactorings Pull Up Properties and Pull Up Dependency where applied. Also, an example of a developer response for commit~\cite{dm1} shown in~\autoref{fig:dm_dedup}, where Pull Up Variable and Pull Up Dependency were applied.


\begin{figure}
    \vspace{-0.3cm}
    \begin{devbox}{Developer Response}
        This was purely to dedup code, so gradle changes could be made in a single place, instead of multiple files, for each sub project.
    \end{devbox}
    \vspace{-0.3cm}
    \caption{Developer Response of Commit~\cite{dm1}}
    \vspace{-0.3cm}
    \label{fig:dm_dedup}    
\end{figure}  

Also, implementing the DRY principle is key in reducing code redundancy, and improving clarity and maintainability while reducing the risk of future bugs. An example of DRY change in commit~\cite{ref11}
where the developer states in the commit message \textit{"Refactor of productFlavors So now we aren't duplicating a bunch of buildConfigField values"}.

\textbf{Security}: This technical debt was mainly addressed by \textit{Externalize Properties} refactoring (1.12\%) .
Build scripts may expose sensitive information or have vulnerabilities that could be exploited. Refactoring techniques such as "Externalize Properties" help address these security risks by reducing potential breaches through improved access control and securing configurations, safeguarding the system from attacks. The goal of this refactoring was also described in the commit message \textit{"Fixes security consideration issues"} for the commit~\cite{ref12}, where sensitive build properties were externalized to a separate file that is then added to .gitignore. 


 

 \begin{table}[ht!]
     \centering
     \begin{tabular}{p{2cm}p{1.25cm}p{4.2cm}}
         \toprule
         \textbf{Technical Debt} & \textbf{Percentage} &\textbf{Refactorings}  \\
         \midrule

         Clarity \& Readability   &  39.63\% & RF, IV, RUC, RC  \\
        
         Extensibility \& Maintainability   &   30.66\% & EV, EP, EM, EV \\
        
         Modularity   & 14.95\% & EM, EPUM, PUV, EPDM, PDD \\
        
         Code Duplication & 14.20\%& PUP, PUM, PUD, EPUV, PUV, DRY \\

         Security  &  1.12\% & EP  \\
        
         \bottomrule
     \end{tabular}
     \vspace{-0.1cm}
     \caption{\buildrefactoring Technical Debt Categories}
     \label{tab:technical-debts}
     \vspace{-0.4cm}
\end{table}

\begin{figure}[h]
\begin{standoutfindings}[Finding 2]
Among the 24 refactoring types identified, 20 are linked to reducing technical debt. We identify which of these refactoring types addresses which of the five specific types of technical debt.
\end{standoutfindings}
\vspace{-0.5cm}
\end{figure}

\subsection{RQ3: \rqThreeContent}
\label{subsec:RQ3}


After the manual labeling of refactoring commits undertaken 
in~\autoref{subsec:Refactoring-Identification}, we created a gold set of labeled commits and corresponding refactorings within them. We use this labeled set to evaluate the performance of \toolName. 
We calculate the precision (PR), recall (RE), and F1-score (F1) of \toolName across the different refactoring types. The results of this evaluation are shown in \autoref{tab:toolperformance}. It's important to note that the refactoring code examples used within the One Shot variant of the prompt were removed from this evaluation set when evaluating the One-Shot variant. This was done to avoid information leakage and ensure the accuracy of our evaluation. 


Using Zero-Shot prompting, \toolName yielded mixed results across the 24 refactoring types, achieving an overall precision of 0.67, recall of 0.72, and F1-score of 0.66. Notably, two refactorings such as Extract and Move Variable and Push Down Variable achieved 
an F1-score of 0.95 and 1 respectively, indicating a strong alignment of \toolName with the empirical definitions in this category. In four refactoring types—Extract and Pull Up Module, Extract Module, Push Down Dependency, and Move Dependency—\toolName maintained a high degree of performance in its classification, with 
an F1-score from 0.7 to 0.84. However, a notable deviation was observed for certain types, including Push Down Module, Extract Method, Move Variable, and Pull Up Properties, which presented lower F-1 scores, between 0.25 and 0.57,
suggesting that the Zero-Shot model struggles with nuances in detecting these refactorings without prior context.

A significant improvement in \toolName performance across several refactoring types can be seen when using One Shot Prompting. 
The overall precision, recall, and F1-score go up to 0.79, 0.78, and 0.75 respectively.
Going over the performance of \toolName over the different refactoring types, 
\toolName demonstrated enhanced performance in detecting 20 out of the 24 refactorings overall.
A notable example is the  dramatic increase in F1-score from  0.25 to a 1 for the Extract and Push Down Module refactoring.
This substantial improvement highlights the effectiveness of providing contextual examples, enabling the model to better interpret and accurately detect code changes without generating false positives or missing instances. Additionally, in two other refactoring categories—Extract Module and Extract and Move Variable—\toolName maintained high precision (1 and 0.96) alongside near-perfect recall, achieving 0.95 in both cases.

However, despite the tool achieving significant recall improvements for Rename Field, Remove Unused Code, Scheduling Tasks, Reformat Code, and Move Variable, ranging from 0.81 to 1, the precision for these categories was noticeably lower, falling between 0.55 and 0.72.
This indicates that, although the tool successfully identified every true instance of these refactorings, it also misclassified some unrelated changes as refactorings, resulting in moderate precision.
Thorough manual validation, it became apparent that some of these refactorings were often intertwined with other refactoring actions, which contributed to the false positives observed. 
\toolName tended to identify them as distinct refactorings even when they were embedded within broader refactoring activities, thus inflating the number of false positives. For example, Move Variable was frequently accompanied by actual refactorings such as Move Method, Move Task, or Extract Module. Remove Unused Code frequently overlaps with refactorings that involve moving code to another class or module (e.g Extract Module, Move Method etc..). 

\begin{figure}[ht!]
   
    \centering
    \fbox{\includegraphics[width=0.95\linewidth]{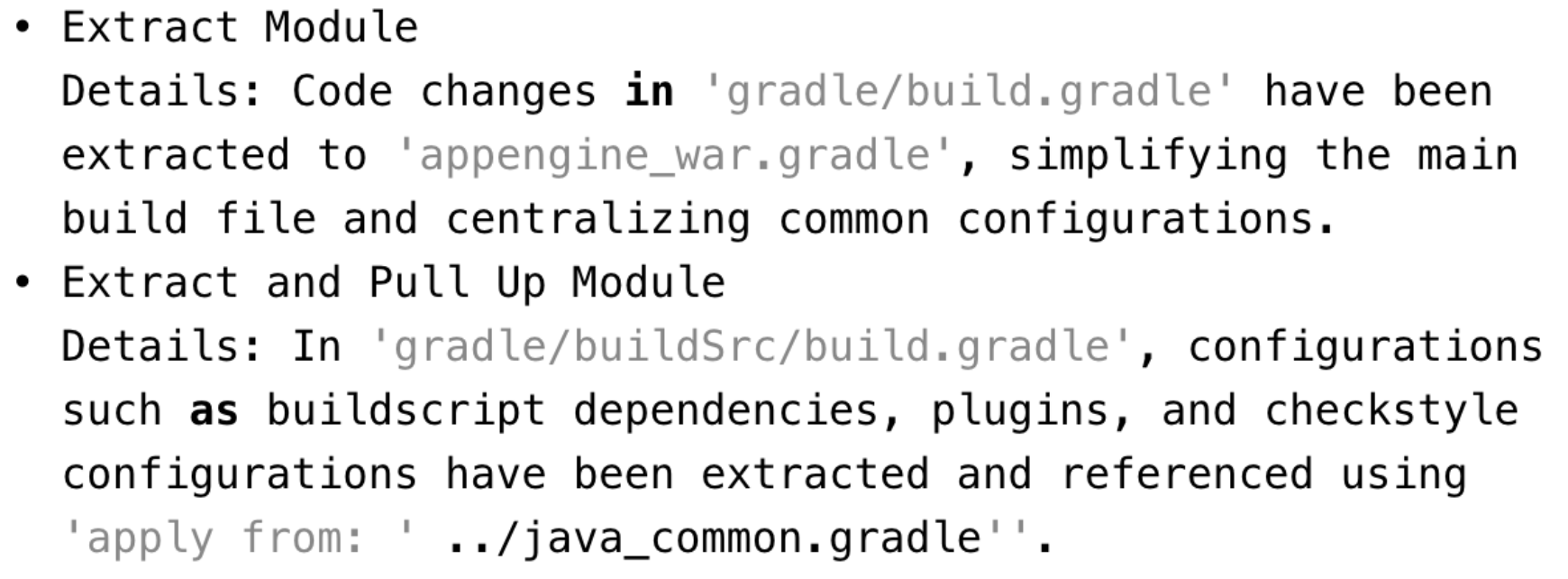}}
    \caption{\toolName output on Commit~\cite{ex1}}
    \label{fig:example}
    \vspace{-0.1cm}
\end{figure}

\begin{table}
\centering
\scriptsize
\begin{adjustbox}{max width=0.7\textwidth}
\begin{tabular}{|>{\centering\arraybackslash}m{3cm}|c|c|c|c|c|c|}
\hline
\multirow{2}{*}{\textbf{Refactoring Types}} & \multicolumn{3}{c|}{\textbf{Zero-Shot}} & \multicolumn{3}{c|}{\textbf{One-Shot}} \\ \cline{2-7}
    & \textbf{PR} & \textbf{RE} & \textbf{F1} & \textbf{PR} & \textbf{RE} & \textbf{F1} \\ 
\hline
Rename Field & 0.54 & 0.71 & 0.61 & 0.72 & 0.82 & 0.76 \\ \hline
Remove Unused Code & 0.53 & 0.77 & 0.63 & 0.56 & 0.81 & 0.66 \\ \hline
Scheduling Tasks & 0.48 & 0.84 & 0.60 & 0.55 & 1.00 & 0.70 \\ \hline
Reformat Code & 0.53 & 0.80 & 0.64 & 0.55 & 0.83 & 0.66 \\ \hline
Extract and Pull Up Module & 1.00 & 0.75 & 0.81 & 1.00 & 0.78 & 0.88 \\ \hline
Extract and Push Down Module & 0.17 & 0.50 & 0.25 & 1.00 & 1.00 & 1.00 \\ \hline
Extract Module & 0.79 & 0.95 & 0.84 & 0.96 & 0.95 & 0.93 \\ \hline
Extract Method & 0.33 & 0.57 & 0.42 & 0.75 & 0.6 & 0.55 \\ \hline
Pull Up Method & 1.00 & 0.50 & 0.67 & 1.00 & 1.00 & 1.00 \\ \hline
Extract Task & 0.50 & 0.63 & 0.55 & 0.58 & 0.63 & 0.60 \\ \hline
Push Down Task & 0.75 & 0.67 & 0.70 & 0.75 & 0.67 & 0.7 \\ \hline
DRY & 0.75 & 0.60 & 0.67 & 0.88 & 0.80 & 0.84 \\ \hline
Push Down Dependency & 0.84 & 0.84 & 0.84 & 0.84 & 0.84 & 0.84 \\ \hline
Pull Up Dependency & 0.82 & 0.53 & 0.64 & 0.95 & 0.53 & 0.67 \\ \hline
Move Dependency & 0.78 & 0.81 & 0.70 & 0.78 & 0.81 & 0.70 \\ \hline
Pull Up Properties & 0.52 & 0.63 & 0.57 & 0.65 & 0.67 & 0.66 \\ \hline
Externalize Properties & 0.73 & 0.57 & 0.64 & 0.81 & 0.65 & 0.72 \\ \hline
Extract Variable & 0.74 & 0.74 & 0.74 & 0.77 & 0.78 & 0.77 \\ \hline
Inline Variable & 0.60 & 0.55 & 0.51 & 0.83 & 0.58 & 0.61 \\ \hline
Move Variable & 0.50 & 0.50 & 0.50 & 0.67 & 1.00 & 0.80 \\ \hline
Pull Up Variable & 0.78 & 0.47 & 0.58 & 0.62 & 0.53 & 0.57 \\ \hline
Push Down Variable & 1.00 & 1.00 & 1.00 & 1.00 & 1.00 & 1.00 \\ \hline
Extract And Move Variable & 1.00 & 0.90 & 0.95 & 1.00 & 0.95 & 0.96 \\ \hline
Extract And Pull Up Variable & 0.50 & 0.62 & 0.49 & 0.83 & 0.68 & 0.68 \\ \hline
\textbf{All Refs.} & \textbf{0.67} & \textbf{0.72} & \textbf{0.66} & \textbf{0.79} & \textbf{0.78} & \textbf{0.76} \\ \hline
\end{tabular}
\end{adjustbox}
\caption{\toolName Performance across the refactoring types with Zero-Shot and One-Shot Prompting.}
\label{tab:toolperformance}
\vspace{-0.3cm}
\end{table}
To demonstrate the usefulness of \toolName on projects in the wild, we share the result of running it on the gradle-modifying commit~\cite{ex1} from \texttt{Google/Nomulus}. This particular commit involved multiple modifications across different build files at various levels of the project hierarchy, making it hard to parse manually. 
\toolName categorized the changes as shown in \autoref{fig:example}, of which we manually verified the veracity, thus giving a glimpse into the usefulness and time-efficiency of our tool for future research.

\begin{figure}
\begin{standoutfindings}[Finding 3]
    We tested two \toolName variants, finding that the One-Shot prompt variant achieved Precision, Recall, and F-1 scores of 0.79, 0.78, and 0.76, respectively, outperforming the Zero-Shot variant, which scored 0.67, 0.72, and 0.66.
\end{standoutfindings}
\end{figure}

\vspace{-0.1cm}
\section{Related Work}
\label{sec:context}

Recent research has highlighted the significance of build code maintenance as a software system evolves. Mcintosh et al.~\cite{mcintosh2014mining} have investigated  the co-change of source and test code with 
build  files.
However, they did not investigate the specific types of build changes, unlike our sharp focus on build refactorings.
Macho et al.~\cite{macho2017extracting} have extracted detailed build changes from Maven build ﬁles. 
However, all of the change types studied are CRUD (create, remove, update, delete) changes to meet new build requirements. Hence, they are not refactorings as they modify the system;s behavior.
In their study, Hardt et al.~\cite{hardt2013ant} have introduced Formiga for Ant, which aims to facilitate build maintenance and dependency discovery. It only includes a limited set of rename and code cleanup refactorings.
Simpson et al.~\cite{duncan2009thoughtworks}, have discussed the need for Ant build files refactoring and lists 23 such refactorings.
However, both these works is focused only on Ant build systems and the taxonomy/categories provided is based on the Author's experience and subjective opinions
contrasting with our research which incorporates a broad empirically-grounded quantitative and qualitative analyses that also includes the more modern Maven and Gradle build systems.
Shridhar et al.~\cite{shridhar2014qualitative} conducted a qualitative analysis of the build commit history of 18 open-source projects from the Maven and Ant systems. 
Their study highlights that changes are predominantly functional, and while 
"Preventive" and "Perfective" improvements are mentioned,  
they are not a focus of this study and the different refactorings that may constitute these changes are not detailed.
In their research, Xiao et al.~\cite{xiao2021characterizing} analyzed 500 commits across 291 projects' Maven build systems, identifying technical debt (TD) primarily arising from tool constraints, library limitations, and dependency management challenges. They also notified developers of TD presence and, in some cases, submitted pull requests to address rudimentary TDs related to dependency incompatibilities.
However, they did not investigate or propose exhaustive and generic refactoring approaches, nor did they link specific fix and TD categories.
Our study proposes a deeper investigation into  build system refactorings and how they relate to technical debt. It builds on previous research by empirically examining various build refactoring changes made by practitioners. We aim to make their findings more accessible and useful by offering a tool, \toolName, which automates the detection of build refactorings to support ongoing research in this field.

\section{Threats To Validity}
\label{sec:threats}
\textbf{Internal Validity:} These threats may stem from errors in the categorization of refactoring changes in build code, which could introduce bias due to limited supporting documentation. To reduce this risk, a panel of three experts independently evaluated selected commits, where they documented refactorings and justifications with caution, only when highly confident. 
Results showed strong agreement across identification and classification.

\textbf{External Validity:} 
Random sampling may have excluded some large commits, potentially missing certain refactorings. However, we applied a stratification strategy to obtain a 95\%-confidence representative sample,  and the commits selected span a large variety of systems, giving credence to our results.

\textcolor{black}{For BuildRefMiner evaluation, as we adopted a stratified sampling, some sub-categories have small class sizes. Therefore, the evaluation metrics may not be representative. Increasing the sample size of each sub-category is ideal but would require more manual labeling. We remedy this with a more broad evaluation of the overall performance of BuildRefMiner across all refactoring types. A more exhaustive evaluation can be part of future work.}

\textbf{Construct Validity:} 

A potential risk is the possibility that commits categorized as refactorings may actually break the build, thereby contradicting the intent of refactoring as behavior-preserving changes. To mitigate this threat, we employed a multi-round, multi-labeler process where all labelers confirmed no behavioral changes occurred in the identified refactorings, ensuring accurate classification.

\section{Study Implications}
\label{sec:Implications}


\textbf{For Researchers:}
We help familiarize researchers with build refactoring, by providing them with an empirically-grounded and definition-supported Taxonomy composed of 24 Build refactorings, and we highlight the 5 TDs some of these refactorings address.
Furthermore, we believe that the 2117 Build Refactoring commits we collected, \toolName which we developed,  and the findings we reach can provide a groundwork to help guide future research into the field of Build refactoring.




\textbf{For Practitioners:}
Due to the lack of empirically-grounded existing taxonomies on Build refactoring, 
practitioners are unaware or misinformed about this practice.
We believe the taxonomy we provided, along with the descriptions of the different TDs addressed by the refactorings within this taxonomy, 
can help guide practitioners in performing Build refactoring procedures. BuildRefMiner is intended for researchers, primarily for automatic large-scale build refactorings mining. We hope to make it more useful for developers in future work via automatic TD identification and remediation.

\section{Conclusion}
\label{sec:conclusion}
In this study, we performed an empirical analysis on \totalCommits commits that were related to build code refactoring from \totalProjects open-source projects. Through manual analysis, we classified the 24 build-related refactorings we discovered into \totalCategories main categories, and we distinguish \totalBuildSpecificRefactoringCategories of which as build-specific.
We also linked some of them to \totalTechnicalDebts TD categories with the help of developer feedback. Finally, we developed \toolName to help guide future research into this field. To the best of our knowledge, this is the first empirical study of refactorings and technical debt in Build Systems.

\section{Acknowledgment}
The UofM-Dearborn authors are supported in part by NSF Award \#2152819.



\bibliographystyle{IEEEtran}
\bibliography{MSR2025}
\end{document}